\documentclass[aps,preprint,showpacs,amsmath,amssymb]{revtex4-1}
\usepackage{graphicx}
\usepackage{dcolumn}
\usepackage{bm}
\usepackage{hyperref}
\usepackage{xcolor}
\usepackage{soul}
\begin{document}

\title{Detection of entanglement via moments of positive maps}
\author{Mazhar Ali \footnote{Email: mazharaliawan@yahoo.com or mazhar.ali@iu.edu.sa}}
\affiliation{Department of Electrical Engineering, Faculty of Engineering, Islamic University of Madinah, 
107 Madinah, Saudi Arabia}


\begin{abstract}
We have reexamined the moments of positive maps and the criterion based on these moments to detect entanglement. For two qubits, we observed that reduction map is equivalent to partial transpose map as the resulting matrices have the same set of eigenvalues although both matrices look different in same computational basis. Consequently, the detection power of both maps is same. For $2 \otimes 4$ systems, we find that moments of reduction map are capable to detect a family of bound entangled states. For qutrit-qutrit systems, we show that moments of reduction map can detect two well known families of bound entangled states. The moments of another positive map can detect the complete range of entanglement for a specific family of quantum states, whereas the earlier criterion fails to detect a small range of entangled states. For three qubits system, we find that applying reduction map to one of the qubit is equivalent to partial transpose operation. In particularly, for GHZ state and W state mixed with white noise, all the moments of a reduction map are exactly the same as the moments of partial transpose map.    
\end{abstract}


\maketitle

\section{Introduction}\label{S-intro}

Quantum entanglement not only defies our intuition about the physical world, but it also has potential utilization for upcoming technologies \cite{Erhard-NRP-2020,Friis-2019}. The theory of entanglement has progressed over the last few decades. The characterization and detection of entanglement is an active area of research and many people have devoted considerable efforts to this field  
\cite{Horodecki-RMP81,Guehne-PR474,Eisert-RMP82}. Despite all this progress, it is still a challenge to determine entanglement properties of a given quantum state. For bipartite quantum systems with $dim \, \mathcal{H}_{AB} \leq 6$, a necessary and sufficient criterion ({\it Peres-Horodecki criterion}) to detect entanglement has been worked out \cite{Peres-PRL77,Horodecki-PLA223}. Namely, this criterion says that for separable states the partial transpose of a density matrix is positive (PPT). For all other higher dimensions of Hilbert space, being PPT does not guarantee separability and quantum states might be entangled. Realignment criterion \cite{Rudolph-QIP4,Chen-QIC3} is able to detect some PPT entangled states, however it fails to detect many other entangled states. The problem of separability is much richer for multipartite quantum systems, where we have fully separable states, biseparable states, fully inseparable states, and genuine entangled states \cite{Schneeloch-PR2020}. Several authors have studied the problem of entanglement detection among multipartite systems 
\cite{Jungnitsch-PRL106,Novo-PRA88,Hofmann-JPA47,Guehne-NJP12,Guehne-PLA375,Bergmann-JPA46,Zhou-NPJ5,Xu-PRA107}.   

Another issue in detection of entangled states in experiments is the quantum state tomography with large number of qubits. It is both hard and exhaustive to perform state tomography for many qubits. It was found that partial transpose moments (PT-moments) can quantify quantum correlations and they are more feasible than the state tomography 
\cite{Paris-LNP-2004,Tran-PRA92-2015,Tran-PRA94-2016,vanEnk-PRL108-2012}. In addition, it was found that PT-moments are related with randomized measurements and few PT-moments are enough to describe separable states 
\cite{Elben-PRA99,Brydges-Sc364,Elben-PRL125,Huang-NP16}. 
Based on moments of partial transpose matrix, two main criteria were worked out to detect bipartite entanglement 
\cite{Yu-PRL127}. First method was to use the lowest Hankel matrix leading to a criterion also defined in 
\cite{Elben-PRL125}. Second method is a necessary and sufficient condition which is based on some moments of partially transposed matrix. 
A set of inequalities, based on Hankel matrices, were also proposed to detect the entanglement of bipartite quantum states
\cite{Neven-npjQI7}.
The local randomized measurements are also proposed to detect multipartite entanglement \cite{Ketterer-PRL122,Ketterer-Q4,Ketterer-PRA106,Ohnemus-PRA107,Wyderka-PRL131,Zhang-arx} and bound entanglement \cite{Imai-PRL126}. The idea of 
PT-moments was extended to realignment matrix for a quantum state. It was shown that the moments of a realigned matrix can detect some bound entanglement \cite{Zhang-QIP21}. 
Recently, an interesting method is proposed to detect entanglement in arbitrary dimensional bipartite quantum systems. The authors proposed to use moments of a Hermitian matrix $(\rho^R)^\dagger \rho^R$, where $\rho^R$ is the well known realigned matrix for a given quantum state. It was shown that this method is capable of detecting entanglement for negative partial transpose states and also some bound entangled states \cite{Aggarwal-PRA109,Aggarwal-PRA108}.  
Recently, the relationship between principle minors and PT-moments is worked out where it was shown that all the PT-moments can be written as a simple function of principle minors 
\cite{Ali-QIP22,Zhang-AP534,Zhang-PRA108}. A more general approach was recently introduced to define the moments of positive maps which are not completely positive \cite{Wang-EPJ137}. Based on moments of positive maps, two criteria were proposed to detect entanglement. As transposition is a positive map which is not completely positive, therefore PT-moments criterion is a special case of this general criterion. Although, the criteria based on moments of positive map were able to detect entanglement for a specific family of quantum states, nevertheless there is some range of parameter where these criteria fail to detect entangled states \cite{Wang-EPJ137}. 

In this work, we propose a set of inequalities resulting from the condition that Hankel matrices are positive for separable states. If the determinant of second order Hankel matrix is negative then the given quantum state is entangled. This proposal is an extension of criterion for partial transpose moments \cite{Neven-npjQI7} to more general positive maps. As a result, we  show that our suggestion is capable of detecting entanglement for several families of quantum states in various dimensions of Hilbert space. In addition, our proposal is capable to detect the complete range of entanglement for a specific family of states where the earlier criterion fails to do so \cite{Wang-EPJ137}. We find out that reduction map for qubits is equivalent to partial transpose map as the resulting matrices give the same set of eigenvalues for both maps. Consequently, the detection of entanglement is optimal for qubit-qubit system. We also extend the idea of moments of positive map to multipartite systems. In multiqubits systems, reduction map applied to anyone of the qubit is equivalent to partial transpose map. For GHZ and W states mixed with white noise, we find that all the resulting moments for reduction map are exactly the same as PT-moments.  

This work is organized as follows. In section \ref{Sec:Model}, we briefly describe positive maps which are not completely positive and discuss the criteria for detection of entanglement. In section \ref{Sec:BP}, we provide many examples of bipartite quantum states detected by our proposal. The extension for multiqubits system is proposed in section \ref{Sec:MQ} and we conclude our work in section \ref{Sec:conc}. 

\section{Positive maps and their moments} 
\label{Sec:Model}

Let $\mathcal{H} = \mathcal{H}_A \otimes \mathcal{H}_B$ be the bipartite finite dimensional Hilbert space. A mixed bipartite quantum 
state $\varrho \in \mathcal{H}$ is a positive semidefinitive density matrix (no negative eigenvalues) and with unit trace. A density matrix is said to be separable (nonentangled) if it can be written as 
\begin{eqnarray}
\varrho^{AB} = \sum_i^n \, p_i \,  \varrho_i^A \otimes \varrho_i^B \, 
\label{Eq:sep}
\end{eqnarray}
where $p_i \geq 0 $ and $\sum_i \, p_i = 1$. In general, it is not easy to verify for a given quantum state whether it is a convex 
combination of product states. To find out whether a given quantum state is entangled or not, is still an open question despite tremendous efforts in this area of research \cite{Horodecki-RMP81,Guehne-PR474}. A necessary condition for separability is to check the partial transpose of the density matrix. If $(\varrho^{AB})^{T_B}$ is negative (having atleast one negative eigenvalue) then state 
$\varrho^{AB}$ is entangled. This condition is necessary and sufficient for $ 2 \otimes 2 $ and $ 2 \otimes 3$ quantum systems. However, 
for higher dimensions of Hilbert space, there are quantum states having positive partial transpose (PPT) nevertheless entangled. 
Recently, partial transpose criterion was reformulated via partial transpose moments (PT-moments) \cite{Elben-PRL125}. This was required to address the issue of state tomography with large number of qubits or higher dimensional systems. The PT-moments are defined \cite{Elben-PRL125} as 
\begin{equation}
p_k = Tr \, [ \, ( \varrho^{AB})^{T_B} )^k \, ] \,, 
\label{Eq:ptm}
\end{equation}
where $k = 1, 2, \ldots , d$ with $d$ as dimension of Hilbert space $\mathcal{H}$. It was found that we do not have to compute all $ d $ - PT-moment, in fact only few PT-moments are sufficient to determine the separability properties of the state 
\cite{Elben-PRL125}. Let $B_k (P)$ are 
the $(k+1) \times (k+1)$ Hankel matrices containing moments $p_k$, such that we have  
\begin{equation}
\big[ \, B_k (P) \big] = p_{i+j+1 } \, , 
\label{Eq:PA}
\end{equation}
where $i, j = 0, 1, \ldots , k$. First two Hankel matrices are given as 
\begin{eqnarray}
B_1 &=& \left[ 
\begin{array}{cc}
p_1 & p_2 \\ 
p_2 & p_3  
\end{array}
\right] \, , \nonumber \\
B_2 &=& \left[ 
\begin{array}{ccc}
p_1 & p_2 & p_3 \\ 
p_2 & p_3 & p_4 \\ 
p_3 & p_4 & p_5 
\end{array}
\right] \, .
\label{Eq:hm}
\end{eqnarray}
A criterion was formulated based on the Hankel matrices that for separable states, Hankel matrices are positive, i.e., $B_k \geq 0$ \cite{Elben-PRL125}. 

The two criteria for detection of entanglement worked out in \cite{Yu-PRL127} can be summarized as follows:

{\bf Result 1}: For a separable state $\rho_{AB}$, we have $B_m (p) \geq 0$ for $m = 1, 2$, etc., and violation means an entangled state for $B_m (p) < 0$. However, it may be that $B_m(p) \geq 0$ for entangled states as well. This criterion is called '$p_3-$ criterion'. 
   
{\bf Result 2}: For a separable state $\rho_{AB}$, the PT-moment $p_3$ satisfies the relation
\begin{eqnarray}
p_3 \geq \, \alpha \, y^3 + (1- \alpha \, y)^3 \,,
\end{eqnarray}
where $y = [\alpha + \sqrt{\alpha ( (\alpha +1) p_2 -1)}\, ] / [\alpha (\alpha +1)]$ with $\alpha = \lfloor 1/p_2 \rfloor$. This criterion is called optimal PT-moment criterion or '$p_3$-OPPT criterion'\cite{Yu-PRL127}. 

In our recent study, we have observed that for the majority of quantum states, only $B_1$ and $ B_2$ are sufficient to determine the separability of a given state. In addition, it is obvious from the formulation of PT-moments that they can detect only the NPT region of quantum states and can not detect PPT-entangled states or even genuine entanglement in multipartite settings \cite{Ali-QIP22}. 
The above method of PT-moments to detect entanglement was extended to define the realignment-moments \cite{Zhang-QIP21}.  Please note that the following method of entanglement detection is different from the recently proposed method in \cite{Aggarwal-PRA109,Aggarwal-PRA108}. The moments for realigned matrix are defined as 
\begin{eqnarray}
r_n = Tr \, [ \, ( R( \rho_{AB} )^n \, ] \, ,
\end{eqnarray}
where $R(\rho)$ is the realigned matrix \cite{Rudolph-QIP4,Chen-QIC3}. As it is known that realignment criterion can detect specific bound entangled states, therefore it is natural to expect that the Hankel matrices based on moments of realigned matrix may detect some PPT-entangled states. However, we should keep in mind that realignment criterion is unable to detect many other entangled states. In \cite{Zhang-QIP21}, the authors gave some examples to support their proposal and they also discussed the corresponding two relations proposed in \cite{Yu-PRL127}.

Recently, the more general approach to this problem was adopted to extend PT-moments to positive-map moments (PM-moments) \cite{Wang-EPJ137}. If $\Lambda$ is a positive map but not completely positive then for a separable state $\rho_{AB}$, the matrix $\mathcal{I} \otimes \Lambda (\rho_{AB})$ must have all positive eigenvalues. This condition is both necessary and sufficient for detection of entanglement \cite{Horodecki-PLA223}. This means that if a given quantum state $\sigma_{AB}$ is entangled then there must exist a positive map $\Lambda$, such that the matrix 
$\mathcal{I} \otimes \Lambda (\sigma_{AB})$ will have at least one negative eigenvalue. 
Transposition is one such positive map, which is not completely positive and hence can detect some entangled states. This means that PT-moments method is only an example for a larger set of positive maps which are not completely positive. Therefore, it follows that the problem of detection of entanglement is to find the positive maps which are not completely positive. This issue is non-trivial because it is not easy to find the positive maps which are not completely positive. Even if we are able to find such maps, it is not clear whether they will detect a given quantum state. Hence the challenge is to look for quantum state specific positive maps as other positive maps may not detect entanglement of a given state. 

The moments of a positive map will be called PM-moments (positive map moments). These are defined as 
\cite{Wang-EPJ137}   
\begin{eqnarray}
q_k = Tr \, [ \, ( \mathcal{I} \otimes \Lambda (\rho_{AB} ) \, )^k \,] \, ,
\label{Eq:mpm}
\end{eqnarray}
where $\Lambda$ is a positive map but not completely positive. This is a more general definition and PT-moments can be regarded as a special case of this general formalism. Hence, in general, if we have some positive maps which are not completely positive, then it is expected that moments of this maps may be able to detect some entangled states including PPT-entangled states. Once, we have the PM-moments, we can write the first two Hankel matrices 
\begin{eqnarray}
S_1 &=& \left[ 
\begin{array}{cc}
q_1 & q_2 \\ 
q_2 & q_3  
\end{array}
\right] \, ,\nonumber \\
S_2 &=& \left[ 
\begin{array}{ccc}
q_1 & q_2 & q_3 \\ 
q_2 & q_3 & q_4 \\ 
q_3 & q_4 & q_5 
\end{array}
\right] \, .
\label{Eq:hm}
\end{eqnarray}
The authors in \cite{Wang-EPJ137} wrote the two general criteria for entanglement detection as below. We note that these two criteria are essentially simple extension of the above criteria by replacing PT with PM as follows:  

{\bf Result } $\tilde{1}$: For a separable state $\rho_{AB}$, we have $S_m (q) \geq 0$ for $m = 1, 2$, etc., and violation means an entangled state for  $S_m (q) < 0$.    

This means that if $S_m(q) < 0$, the state is entangled, however, it may be that $S_m(q) \geq 0$ for entangled states as well. The key difference with earlier criterion 1, mention above is the fact that in general the operation $\mathcal{I} \otimes \Lambda (\rho_{AB} )$ may not be trace preserving ({Transposition (including partial transpose) map preserves the trace of a matrix and we always have} $p_1 = 1$). Therefore, we may normalize the mapped matrix by $(\mathcal{I} \otimes \Lambda (\rho_{AB} ))/Tr(\mathcal{I} \otimes \Lambda (\rho_{AB} ))$ with the consequence that $q_1 = 1$, otherwise $q_1 \neq 1$. The normalization process is not really required as the positivity or negativity of a matrix does not depend on it. For a normalized matrix, the criterion proposed in Ref.\cite{Wang-EPJ137} is $q_3 - q_2^2 \geq 0$ for any positive map and for separable states. This criterion is called '$q_3-\Lambda$' criterion. 
   
{\bf Result } $\tilde{2}$: For a separable state $\rho_{AB}$, the PM-moment $q_3$ satisfies the relation
\begin{eqnarray}
q_3 \geq \, \beta \, x^3 + (1- \beta \, x)^3 \,,
\end{eqnarray}
where $x = [\beta + \sqrt{\beta ( (\beta +1) q_2 -1)}\, ] / [\beta (\beta +1)]$ with $\beta = \lfloor 1/q_2 \rfloor$. This criterion is called optimal PM-moment criterion or '$q_3-O\Lambda$'. This is essentially the same result as above mentioned $p_3$-OPPT criterion for positive maps.

Now we are in a position to present our main contribution in this article. The main motivation behind this work is the unability of earlier proposal to detect some entangled states. As an example, in Ref.\cite{Wang-EPJ137}, the authors proposed '$q_3-\Lambda$' and '$q_3-O\Lambda$' as two criteria to detect entangled states. However, in their work, the authors provided a family of entangled states and both of these criteria were unable to detect the complete range of entanglement for the given quantum states. We will provide details for this situation in subsection below, where we discuss entanglement detection for $3 \otimes 3$ systems. 

{\bf Proposal }: Our proposal is in the similar lines to Ref. \cite{Neven-npjQI7}. We propose to use the Hankel matrices to detect entanglement or alternatively to check a set of inequalities resulting from determinant of the Hankel matrices $S_m (q)$ for detection of entanglement such that $S_m (q)< 0$ means entangled states with appropriate positive map for a given state. Because, as it is mentioned before, a different positive map may not be able to detect entanglement of a given state. For most of the cases, $S_1 (q)$ and $S_2(q)$ would detect the complete range of entanglement. If needed, one can add more moments and check $S_3(q)$. 

{\bf Theorem} : {\it For a separable state the Hankel matrices } $S_m(q)$ {\it are positive and if any such matrix is negative then the state is entangled}. \newline
{\it Proof}: The proof of this theorem is very simple. For a separable state defined in Eq. (\ref{Eq:sep}) and for a positive map which is not completely positive, the evolved state $\varrho(t) = \mathcal{I} \otimes \Lambda (\varrho^{AB}) = \sum_i^n \, p_i \,  \varrho_i^A \otimes \Lambda (\varrho_i^B) \,$, remains separable and therefore $S_m(q) > 0$. Whereas for any other initial state $\rho^{AB}(t) = \mathcal{I} \otimes \Lambda (\rho^{AB})$, if $S_m(q) < 0$ then state is gauranteed to be entangled as dictated by Horodecki's theorem on separability \cite{Horodecki-PLA223}.      

{\bf Lemma } : {\it For a normalized mapped matrix and for separable states, we have}
\begin{eqnarray}
det \, \left( 
\begin{array}{ccc}
1 & q_{2} & q_{3} \\ 
q_2 & q_{3} & q_{4}\\
q_3 & q_4 & q_5  
\end{array} \right) \, \geq 0 \,.
\label{Eq:br}
\end{eqnarray}
This Lemma is equivalent to say that the Hankel matrix $S_2(q)$ is positive for a separable state. An alternative procedure to verify the positivity of a matrix is via principal minor (PM), which is defined as the determinant of a sub-matrix obtained by removing some rows and columns from the main matrix. We denote the sub-matrices with symbols $S(i)$, $S(ij)$, and $S(ijk)$ and their determinants with symbols $[ S(i) ]$, $[ S(ij) ]$, and $[ S(ijk) ]$. A result in matrix theory says that if a matrix is positive semidefinite then all its principal minors are also positive semidefinite \cite{Horne}. The positivity of Eq.\ref{Eq:br} implies that the following set of inequalities must be satisfied by separable states.
\begin{eqnarray}
q_3 \geq 0 \,,\nonumber \\ 
q_5 \geq 0 \,,\nonumber \\ 
q_3 \geq q_2^2 \,,\nonumber \\
q_3 \, q_5 \geq q_4^2 \,, \nonumber \\
q_5 \geq q_3^2 \,,\nonumber \\
q_3 \, q_5 - q_4^2 - q_2^2\, q_5 + 2 q_2 \, q_3 \, q_4 - q_3^3 \geq 0 \, .
\label{Eq:Inq}
\end{eqnarray}
The violation of any of this condition is equivalent to say that matrix $S_2(q)$ is negative and hence the respective quantum state is entangled. In next sections, we will provide several examples to show that this criterion is able to detect entangled states whereas earlier criteria fails to do so. Hence our proposal is an enhancement and improvement in this direction of entanglement detection. 

\section{Bipartite quantum systems}
\label{Sec:BP}

In this section, we provide several examples of bipartite quantum states detected by criterion based on at least first two Hankel matrices. We also discuss some positive maps which are not completely positive and their detection capability in each of subsections below.   
 
\subsection{Qubit-Qubit systems} 

This is the simplest bipartite quantum system. The separability problem for this dimension of Hilbert space is completely solved. Namely, all quantum states whose partial transpose w.r.t any subsystem is again a density matrix, are separable. This system is also unique in the sense that several measures of entanglement have their explicit analytical expressions available \cite{Horodecki-RMP81,Guehne-PR474}. Let 
$E_{ij} = |i\rangle\langle j|$ be an elementary operator mapping an orthonormal state $|j\rangle$ to an orthonormal state $|i\rangle$. Other than transposition, two positive maps which are not completely positive \cite{Hou-2010,Hou-2011} can be written in terms of elementary operators as 
\begin{eqnarray}
\Lambda_1 (A) =  E_{11} \, A \, E_{11}^\dagger + E_{22} \, A \, E_{22}^\dagger + E_{12} \, A \, E_{12}^\dagger \nonumber \\
+ E_{21} \, A \, E_{21}^\dagger - (E_{11} + E_{22}) A (E_{11} + E_{22})^\dagger \, , 
\label{Eq:mp1}
\end{eqnarray}
and 
\begin{eqnarray}
\Lambda_2 (A) =  (2 \, E_{11} + E_{22} )\, A \, (2 \, E_{11} + E_{22} )^\dagger + E_{12} \, A \, E_{12}^\dagger  \nonumber \\
+ E_{21} \, A \, E_{21}^\dagger  - (E_{11} + E_{22}) A (E_{11} + E_{22})^\dagger \, . 
\label{Eq:mp2}
\end{eqnarray}
Both maps are positive because they map positive matrices to positive matrices as
\begin{eqnarray}
\Lambda_1 \, \left( 
\begin{array}{cc}
a_{11} & a_{12} \\ 
a_{21} & a_{22}  
\end{array}
\right) =  \left( 
\begin{array}{cc}
a_{22} & -a_{12} \\ 
-a_{21} & a_{11}  
\end{array}
\right) \,,
\label{Eq:mp12}
\end{eqnarray}
and 
\begin{eqnarray}
\Lambda_2 \, \left( 
\begin{array}{cc}
a_{11} & a_{12} \\ 
a_{21} & a_{22}  
\end{array}
\right) =  \left( 
\begin{array}{cc}
3 \, a_{11} + a_{22} & a_{12} \\ 
a_{21} & a_{11}  
\end{array}
\right) \,.
\label{Eq:mp22}
\end{eqnarray}
Note that the map $\Lambda_1$ (this map is called the reduction map) in Eq.(\ref{Eq:mp1}) swaps the diagonal elements and introduces negative signs to off-diagonal elements, whereas $\Lambda_2$ swaps the diagonal elements with addition of $3 a_{11}$ with $a_{22}$. We will show below that due to this reason, $\Lambda_2 $ is not a good map to detect entanglement whereas $\Lambda_1$ is the best as it detects all entangled states for qubit-qubit systems and some for three qubits as described in next section. 

The map $\Lambda_1$ in Eq.(\ref{Eq:mp1}) when applied to qubit B of a qubit-qubit state, transforms the quantum state as
\begin{eqnarray}
\mathcal{I} \otimes \Lambda_1 \, \left( 
\begin{array}{cccc}
\rho_{11} & \rho_{12} & \rho_{13} & \rho_{14}\\ 
\rho_{21} & \rho_{22} & \rho_{23} & \rho_{24}\\ 
\rho_{31} & \rho_{32} & \rho_{33} & \rho_{34}\\ 
\rho_{41} & \rho_{42} & \rho_{43} & \rho_{44}\end{array}
\right) =  \left( 
\begin{array}{cccc}
\rho_{22} & -\rho_{12} & \rho_{24} & -\rho_{14}\\ 
-\rho_{21} & \rho_{11} & -\rho_{23} & \rho_{13}\\ 
\rho_{42} & -\rho_{32} & \rho_{44} & -\rho_{34}\\ 
-\rho_{41} & \rho_{31} & -\rho_{43} & \rho_{33}\end{array}
\right) \,.
\label{Eq:mp2qb}
\end{eqnarray}
We note atleast two important changes in the matrix. First, the off-diagonal elements $\rho_{14}$ and $\rho_{23}$ (they are the major contributors in entanglement of qubit-qubit quantum states) are not moved from their positions. Second, the diagonal elements are relocated and the trace is preserved. This operation is in fact equivalent to partial transpose operation where the diagonal elements are fixed in their positions but the elements $\rho_{14}$ and $\rho_{23}$ are swapped. Although the resulting matrices after partial transpose operation and after applying the map $\mathcal{I} \otimes \Lambda_1$ look different in the same computational bases, nevertheless both matrices have the same set of moments. Hence detection of entanglement by PT-moments and moments of map $\Lambda_1$ is exactly same. In this sense they are equivalent maps. 

\begin{figure}[h]
\centering
\scalebox{2.20}{\includegraphics[width=1.99in]{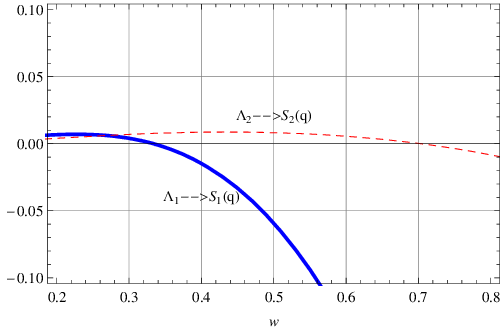}}
\caption{(Color online) Entanglement is optimally detected by map $\Lambda_1$ for $w > 1/3$. The other map $\Lambda_2$ is not optimal and only second Hankel matrix detects some entanglement. See text for more descriptions.}
\label{FIG:22}
\end{figure}

{\bf Example 1}: As a concrete example, let us consider Werner states of two qubits, defined as 
\begin{equation}
\rho_w = w \, |\phi \rangle\langle \phi | + \frac{1 - w}{4} \, \mathbb{I}_4 \,,
\end{equation}     
where $0 \leq w \leq 1$, and $|\phi\rangle = (|00\rangle + |11\rangle)/\sqrt{2}$ is the maximally entangled Bell state. These states are entangled for $w > 1/3$. In Figure \ref{FIG:22}, we plot the minimum eigenvalues for Hankel matrices resulting from positive maps 
$\Lambda_1$ (Eq.\ref{Eq:mp1}) and $\Lambda_2$ (Eq.\ref{Eq:mp2}). We observe that the first Hankel matrix $S_1 (q)$ for positive map 
$\Lambda_1$ becomes negative $S_1 (q) < 0$ precisely for $w > 1/3$. The first Hankel matrix related with second positive map 
$\Lambda_2$ is positive and can not detect any entanglement. However, the second Hankel matrix for positive map $\Lambda_2$ becomes negative for $w > 0.7$, hence detecting some entanglement but not in an optimal manner. 

We have checked for several two-qubit entangled states that $S_1 (q)$ related with $\Lambda_1$ is always negative whenever 
$B_1 (p)$ is negative and vice versa. The map $\Lambda_2$ can not detect all entangled states even with $S_2 (q)$. Therefore, we conclude this subsection with observation that $\Lambda_1$ is the optimal map in detecting entanglement of $2 \otimes 2$ quantum systems and it is equivalent to partial transpose map. 
  
\subsection{$2 \otimes 4$ system} 

This quantum system is the simplest bipartite system with the possibility of having PPT-entangled states. 

{\bf Example 2}: The well known bound entangled states for $2 \otimes 4$ are given as
\begin{eqnarray}
\sigma_b = \frac{1}{7 b+1} \, \left[ 
\begin{array}{cccccccc}
b & 0 & 0 & 0 & 0 & b & 0 & 0 \\ 
0 & b & 0 & 0 & 0 & 0 & b & 0\\ 
0 & 0 & b & 0 & 0 & 0 & 0 & b\\ 
0 & 0 & 0 & b & 0 & 0 & 0 & 0\\
0 & 0 & 0 & 0 & \frac{1+b}{2} & 0 & 0 & \frac{\sqrt{1-b^2}}{2}\\
b & 0 & 0 & 0 & 0 & b & 0 & 0\\
0 & b & 0 & 0 & 0 & 0 & b & 0\\
0 & 0 & b & 0 & \frac{\sqrt{1-b^2}}{2} & 0 & 0 & \frac{1+b}{2}
\end{array}
\right] \,,
\label{Eq:bes1}
\end{eqnarray}
where $0 < b < 1$. The partial transpose of this matrix w.r.t A or B is positive (PPT), however $\sigma_b$ is entangled for $0 < b < 1$\cite{Horodecki-PLA223}. It follows that PT-moments can not detect them. In order to detect them via Eq.(\ref{Eq:hm}), we first need a positive map which is not completely positive. Strictly speaking, we need a map 
$\Lambda_b$: $\mathcal{B}(\mathcal{C}^4) \to \mathcal{B}(\mathcal{C}^2) $ such that 
$\mathcal{I}_2 \otimes \Lambda_b (\sigma_b) < 0$. Unfortunately, we could not find such a map despite some efforts. It is known that finding this map may not be easy. However, there exists a positive map (reduction map) $\Lambda$: 
$\mathcal{B}(\mathcal{C}^4) \to \mathcal{B}(\mathcal{C}^4)$, which is not completely positive \cite{Hou-2010}. We show below that moments of this map can detect entanglement for this state. 

Let $| e_j\rangle$ with $j = 1, 2, 3, 4$, be an orthonormal basis in $\mathcal{C}^4$. We define the elementary operators by 
$E_{ij} = |e_i\rangle\langle e_j|$ (we have $16$ such operators). In addition, we define two operators $F_{ij} = (E_{ii}+E_{jj})/\sqrt{2}$ 
and $G_{ij} = (E_{ii}-E_{jj})/\sqrt{2}$. Then the reduction map can be written as
\begin{eqnarray}
\Lambda (X) = \sum_{i \neq j} \, E_{ij} X E_{ij}^\dagger + \sum_{i \neq j} \, G_{ij} X G_{ij}^\dagger - \sum_{i \neq j} \, 
F_{ij} X F_{ij}^\dagger \,,
\label{Eq:rm}
\end{eqnarray} 
where $X$ is a positive matrix. This map is capable to detect entanglement of bound entangled states defined in Eq.(\ref{Eq:bes1}).   
  
\begin{figure}[h]
\centering
\scalebox{2.28}{\includegraphics[width=1.99in]{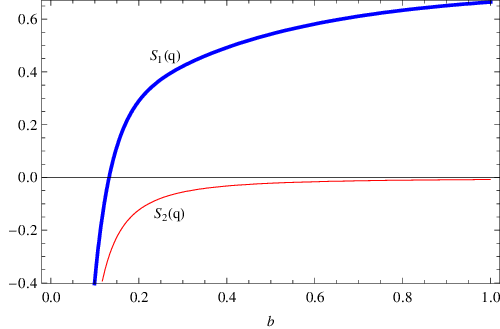}}
\caption{(Color online) Minimum eigenvalues of Hankel matrices $S_1(q)$ and $S_2(q)$ are plotted against parameter $b$. First matrix fails to detect entanglement after $b \approx 0.12$ shown by thick (blue) curve, whereas the matrix $S_2(q)$ is negative for $0 < b < 1$. See text for more details.}
\label{FIG:24}
\end{figure}
In Figure (\ref{FIG:24}), we plot the minimum eigenvalues for the Hankel matrices $S_1(q)$ and $S_2(q)$ against the parameter $b$. We observe that first three moments for the positive map can not completely detect entanglement, whereas $S_2(q)$ is negative for entire range 
$0 < b < 1 $. 

\subsection{$3 \otimes 3$ system} 

It is known that qutrit-qutrit systems have PPT-entangled states as well as NPT and separable states. There are several families of bound entangled states living in this dimension of Hilbert space. It is clear by now that PT-moments are unable to detect bound entangled states. However, we will demonstrate that moments of positive maps are very effective to detect entangled states including PPT-entangled states. 

First we define a positive map $\Phi_1$ which is not completely positive. This map takes operators from $\mathcal{H}_A$ to 
$\mathcal{H}_B$.  The map is defined as \cite{Hou-2011}
\begin{eqnarray}
\Phi_1 (A) = 2 \, \sum_{i=1}^{3} \, E_{ii} \, A \, E_{ii}^\dagger + E_{12} \, A \, E_{12}^\dagger 
+ E_{23} \, A \, E_{23}^\dagger \nonumber \\ +  E_{31} \, A \, E_{31}^\dagger - \bigg(\sum_{i=1}^3 E_{ii} \bigg) \, A \, 
\bigg(\sum_{i=1}^3 E_{ii} \bigg)^\dagger \, . 
\label{Eq:mp3}
\end{eqnarray}
This map can not detect all entangled states, but it is very good to detect entanglement of a specific family of quantum states defined in example $3$.

{\bf Example 3:} Let us consider a specific family of quantum states for $3 \otimes 3$, which are defined as \cite{Horodecki-PRL82}
\begin{eqnarray}
\rho_\alpha = \frac{2}{7} \, |\Psi\rangle \langle \Psi| + \frac{\alpha}{7} \, \sigma_+ + \frac{5 - \alpha}{7} \, \sigma_- \,,
\label{Eq:bes331}
\end{eqnarray}  
where $|\Psi\rangle = (|11\rangle + |22\rangle + |33\rangle)/\sqrt{3}$ is maximally entangled state, 
$\sigma_+ = (|12\rangle\langle 12| + |23\rangle\langle 23| +|31\rangle\langle 31|)/3$, and  
$\sigma_- = (|21\rangle\langle 21| + |32\rangle\langle 32| +|13\rangle\langle 13|)/3$ are separable states. The parameter $\alpha $ has a range $2 \leq \alpha \leq 5$. It was shown that these states are separable for $2 \leq \alpha < 3$, bound entangled for $3 \leq \alpha \leq 4$, and free entangled (having negative partial transpose) for $4 < \alpha \leq 5$ \cite{Horodecki-PRL82}. 
Recently, detection of entanglement by moments of positive maps for these states is studied \cite{Wang-EPJ137}, where the authors showed that $q_3-\Phi$ criterion detects entanglement for $\alpha \in [3.1658, 5]$. The $q_3-O\Phi$ criterion detects entanglement for 
$\alpha \in [3.0291, 5]$. The PT-moments optimal criterion only detects entanglement for $\alpha \in [4.7259, 5]$. This clearly demonstrate that $q_3-O\Phi$ is more powerful criterion to detect entanglement for these quantum states. However, strictly speaking, 
all of these criteria fail to detect entanglement for the range $\alpha \in [3, 3.0291]$ where we already know that the quantum states are entangled. We show below that with our proposal, we are able to detect the complete range of these entangled states defined in Eq.(\ref{Eq:bes331}). 

\begin{figure}[h]
\centering
\scalebox{2.30}{\includegraphics[width=1.99in]{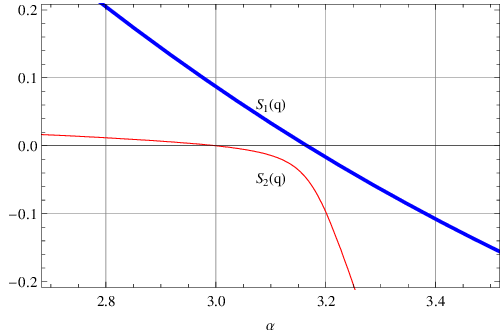}}
\caption{(Color online) Minimum eigenvalues for Hankel matrices are plotted against parameter $\alpha$. The matrix $S_1(q)$ is negative for $\alpha > 3.1658$, whereas $S_2(q)$ is able to detect entanglement for $3 < \alpha \leq 5$. }
\label{FIG:331}
\end{figure}
In Figure (\ref{FIG:331}), we plot the minimum eigenvalues for matrices $S_1(q)$ and $S_2(q)$. The matrix $S_1(q)$ becomes negative for $\alpha > 3.1658$ shown by thick (blue) curve, whereas the matrix $S_2(q)$ becomes negative for $3 < \alpha \leq 5$, detecting the complete range of entangled states.   

{\bf Example 4:} Let us consider another family of bound entangled states for $3 \otimes 3$ system \cite{Horodecki-PLA223}. These states are defined as 
\begin{eqnarray}
\sigma_a = \frac{1}{8 a+1} \, \left[ 
\begin{array}{ccccccccc}
a & 0 & 0 & 0 & a & 0 & 0 & 0 & a \\ 
0 & a & 0 & 0 & 0 & 0 & 0 & 0 & 0\\ 
0 & 0 & a & 0 & 0 & 0 & 0 & 0 & 0\\ 
0 & 0 & 0 & a & 0 & 0 & 0 & 0 & 0\\
a & 0 & 0 & 0 & a & 0 & 0 & 0 & a \\ 
0 & 0 & 0 & 0 & 0 & a & 0 & 0 & 0\\
0 & 0 & 0 & 0 & 0 & 0 & \frac{1+a}{2} & 0 & \frac{\sqrt{1-a^2}}{2}\\
0 & 0 & 0 & 0 & 0 & 0 & 0 & a & 0\\
a & 0 & 0 & 0 & a & 0 & \frac{\sqrt{1-a^2}}{2} & 0 & \frac{1+a}{2}
\end{array}
\right] \,,
\label{Eq:bes2}
\end{eqnarray}
where $0 < a < 1$. It is not difficult to find that partial transpose of these states w.r.t any subsystem is positive, nevertheless the states are entangled \cite{Horodecki-PLA223}. We have observed that the map $\Phi_1$ defined in Eq.(\ref{Eq:mp3}) is unable to detect entanglement of these states. However, the reduction map defined in Eq.(\ref{Eq:rm}) is capable of detecting the complete range of these bound entangled states. We only need to write that map for qutrits as it was defined for operators in Hilbert space with dimension $4$. This process is fairly simple as for qutrits, we have $9$ elementary operators $E_{ij}$, $6$ operators for each $G_{ij}$ and $F_{ij}$ with $i \neq j$. Therefore, it is straightforward to write that map for qutrits. 

\begin{figure}[h]
\centering
\scalebox{2.30}{\includegraphics[width=1.99in]{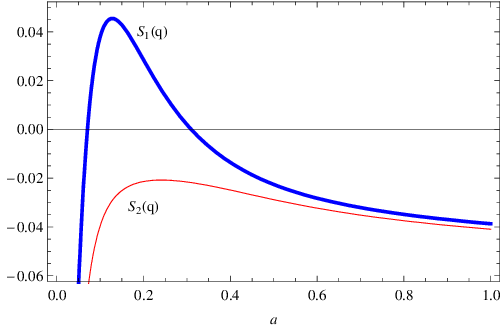}}
\caption{(Color online) Minimum eigenvalues for matrices $S_1(q)$ (thick blue curve) and $S_2(q)$ (red curve) are plotted against 
parameter $a$. See text for details.}
\label{FIG:332}
\end{figure}
Figure (\ref{FIG:332}) depicts the minimum eigenvalues plotted against parameter $a$, for matrices $S_1(q)$ and $S_2(q)$. 
We observe that matrix $S_1(q)$ is able to detect entanglement except for some range, shown by thick (blue) curve. The matrix $S_2(q)$ is negative for the whole range of parameter $0 < a < 1$, detecting entanglement.  

{\bf Example 5}: As our last example for $3 \otimes 3$ systems, we consider the vectors \cite{Bennet-PRL82} 
\begin{eqnarray}
|\psi_0\rangle = \frac{1}{\sqrt{2}} |1\rangle \otimes (|1\rangle - |2\rangle) \, ,  
|\psi_1\rangle = \frac{1}{\sqrt{2}}  (|1\rangle - |2\rangle) \otimes |3\rangle  \,, \nonumber \\
|\psi_2\rangle = \frac{1}{\sqrt{2}} |3\rangle \otimes (|2\rangle - |3\rangle) \,,
|\psi_3\rangle = \frac{1}{\sqrt{2}} (|2\rangle - |3\rangle)  \otimes |1\rangle \,, \nonumber \\
|\psi_4\rangle = \frac{1}{3} (|1\rangle + |2\rangle + |3\rangle)(\langle 1 | + \langle 2 | + \langle 3| ) \,.
\end{eqnarray}
These vectors constitute the set of so called {\it unextendible product basis} (UPB). It follows that the mixed state
\begin{eqnarray}
\varrho = \frac{1}{4} \big[ \, \mathbb{I} - \sum_{x = 0}^4 \, |\psi_x\rangle\langle \psi_x| \, \big]
\end{eqnarray}
is PPT as well as entangled. The reduction map for qutrits is able to detect these states as the minimum eigenvalue for matrix $S_1(q)$ is equal to $- 9/[4(301+\sqrt{91177})]$.
 
\section{Three qubits system}
\label{Sec:MQ}

Let us consider three qubits, $A$, $B$, and $C$. The separability problem for multipartite quantum systems is much richer than bipartite systems. Here we have several classes of quantum states. A quantum state is called fully separable if it can be written as convex combination of product vectors of each qubit, i.e., $\rho = \sum_i p_i \rho_A \otimes \rho_B \otimes \rho_C$. A quantum state can be biseparable if the joint state of three qubits factors into $AB$ and $C$ as an example and their mixtures. Biseparable states can be entangled between two parties and not between three parties. Then there are fully inseparable states and finally genuine multipartite entangled states. For more details see Figure 1 in Ref. \cite{Schneeloch-PR2020}.

In our recent study, we have analyzed the PT-moments for three qubits system and have found that PT-moments are only able to detect the NPT range of quantum states \cite{Ali-QIP22}. This implies that PPT entangled states for three qubits can not be detected by PT-moments. Let us consider that we apply a positive map $\Lambda_c$ which is not completely positive, on qubit C as 
\begin{eqnarray}
\tilde{\rho}_C = \mathbb{I}_2 \otimes \mathbb{I}_2 \otimes \Lambda_c (\rho_{ABC}) \,,   
 \end{eqnarray}
where $\tilde{\rho}_C$ may not have unit trace. We can find the other matrices $\tilde{\rho}_B$ and $\tilde{\rho}_A$ by applying the positive maps to parties $B$ and $A$, respectively. We define the moments of this map as follows:
\begin{eqnarray}
r_n^A \, = \, Tr \, [ \, (\tilde{\rho}_A )^n \, ] \,, \nonumber \\
r_n^B \, = \, Tr \, [ \, (\tilde{\rho}_B )^n \, ] \,, \nonumber \\
r_n^C \, = \, Tr \, [ \, (\tilde{\rho}_C )^n \, ] \,, 
\end{eqnarray}
where $n = 1, 2, \ldots, d$ with $d$ as dimension of Hilbert space. Finally, we define the moments of a positive map as
\begin{eqnarray}
r_n = (r_n^A \, r_n^B \, r_n^C)^{1/3}\,.
\end{eqnarray}
There are always the situations where $r_n^A = r_n^B = r_n^C$ due to symmetry of quantum states. In those cases, it is sufficient to take moments due to one partition. In any case, we can construct the respective Hankel matrices $S_n(r)$  and check the positivity/negativity of these matrices. 

{\bf Example 6}: For three qubits, we take the $GHZ$ state mixed with white noise, given as
\begin{eqnarray}
\rho_{GHZ} = \gamma \frac{\mathbb{I}}{8} + (1-\gamma) |GHZ\rangle\langle GHZ|  \,,
\end{eqnarray}
where $0 \leq \gamma \leq 1$ and $|GHZ \rangle = 1/\sqrt{2} (|000 \rangle + |111 \rangle)$. We apply the reduction map defined in 
Eq.\ref{Eq:mp1} to qubit $C$. We have found out that all the moments of this positive map are exactly the same as the moments for partial transpose map. We do not provide those moments as they are mentioned in Ref.\cite{Ali-QIP22}. Therefore, the map defined in Eq.(\ref{Eq:mp1}) is capable to detect the complete range of $\rho_{GHZ}$ as being NPT.  
 
{\bf Example 7}: The final example is $W$ state mixed with white noise as 
\begin{eqnarray}
\rho_W = \kappa \frac{\mathbb{I}}{8} + (1-\kappa ) |W\rangle\langle W|  \,,
\end{eqnarray}
where $0 \leq \kappa \leq 1$ and $|W \rangle = 1/\sqrt{3} (|001 \rangle + |010 \rangle + |100 \rangle)$. Once again, we find that all the moments of positive map defined in Eq.(\ref{Eq:mp1}) are exactly the same as the moments for partial transpose map. These moments are provided in Ref.\cite{Ali-QIP22}. Hence, this map is capable to detect the NPT region of these states.      

\section{Discussion and Summary} 
\label{Sec:conc}

We studied detection of bipartite and multipartite entanglement via moments of positive maps. Positive maps which are not completely positive, can detect entangled states. However, this problem is not trivial because it is known that finding such maps is not an easy task. Even if we find such maps, it is not clear which entangled states they will optimally detect. It might be that some positive maps are specific to specific quantum states and/or some specific quantum states might have specific positive maps designed to detect them. We proposed a set of inequalities such that if any one of this inequality is violated then the given quantum state must be entangled. This proposal is an extension of partial transpose map inequalities and an improvement to the earlier criteria. We showed that our proposal is capable of detecting the complete range of entanglemnet for a specific family of states, whereas the earlier criteria failed to do so for some range of the parameter. We provided several examples of bipartite quantum states which are completely detected by our proposal. We found that reduction map is equivalent to partial transpose map only for qubit-qubit system and for multiqubit system where map is applied to anyone of the qubit. It is obvious that neither PT-moments nor reduction map can detect PPT entangled states for three qubits. It remains to find the positive maps, which are capable to detect entanglement of PPT entangled states of three qubits. For higher dimensional systems reduction map is not equivalent to partial transpose map. Recently, an interesting proposal investigated the detection of entanglement for bipartite quantum systems. The idea is to use moments of a Hermitian matrix $(\rho^R)^\dagger \rho^R$, where $\rho^R$ is the well known realigned matrix \cite{Aggarwal-PRA109,Aggarwal-PRA108}. We have done some initial investigations whether this proposal can also detect the quantum states detected by our proposal. Our studies suggest that this more recent proposal is good to detect some specific quantum states, however it did not detected the quantum states which are detected by our proposal. It still remains to investigate whether using realignment criterion in this way or some other function of it, can be useful to detect every bound entangled state for bipartite as well as multipartite quantum state. We aim to address these issues in our future work. 
\subsection*{Acknowledgments}
The author is grateful to both reviewers for their constructive comments, useful suggestions and pointing to important references. The author appreciates the additional editorial evaluations.  
%
\subsection*{Author Contributions}
I declare that this is my own work, thought, worked out and compiled all by myself.

\subsection*{Funding}
Not Applicable

\subsection*{Data Availability}
The author can provide any data and material upon request to relevant party.

\subsection*{Code Availability}
The author can provide source code used to compile the results upon request to relevant party.

\section*{Declarations}

\subsection*{Conflict of Interest}
The author declare that they have no conflict of interest.

\subsection*{Ethics approval}
Not Applicable

\subsection*{Consent to participate}
Not Applicable

\subsection*{Consent to publications}
Not Applicable


\end{document}